\documentstyle[12pt]{article}

\advance\textwidth 1cm
\advance\oddsidemargin -0.5cm

\newcommand\beq{\begin{equation}}
\newcommand\eeq{\end{equation}}
\def\beqa{\begin{eqnarray}}
\def\eeqa{\end{eqnarray}}

\def\sHa{{\cal H}}

\def\non{{\nonumber}}
\def\l{{\langle}}
\def\r{{\rangle}}
\def\w{{\omega}}
\def\U{{\cal U}}

\def\vb{{\vphantom{b}}}

\def\d{{\partial}}
\def\dag#1{{\dagger_#1}}

\begin{document}
\baselineskip=24pt

\title{Quantum Canonical Transformations: Physical Equivalence of
Quantum Theories}
\author{Arlen Anderson\thanks{arley@ic.ac.uk}\\
Blackett Laboratory\\ Imperial College\\ Prince Consort
Road\\ London SW7 2BZ England}
\date{February 14, 1993}
\maketitle

\vspace{-10cm}
\hfill Imperial/TP/92-93/20

\hfill hep-th/9302062
\vspace{10cm}

\begin{abstract}
Two quantum theories are physically equivalent if they are related, not by a
unitary transformation, but by an isometric transformation.  The
conditions under which a quantum canonical transformation is an isometric
transformation are given.
\end{abstract}
\newpage

One of the most powerful ways of solving a quantum theory is to make a
canonical
transformation to a simpler theory in different variables.  Following
Dirac\cite{Dir,Mos}, there is a widespread belief that the unitary
transformations are the analog of the classical canonical transformations
in quantum theory.  Quantum canonical transformations can however be
defined without mentioning a Hilbert space structure, and they are in
themselves neither unitary nor non-unitary.  Furthermore,
one finds that non-unitary transformations
play an important role in the integrability of a theory\cite{And}.

Physically, one is often interested not only in solving a theory, but
in finding when it is physically equivalent to another theory expressed
in different variables.  Two theories are physically equivalent if there
is an isomorphism between their states which preserves the values of
inner products, so that all physical amplitudes are the same.  Certainly, two
theories related by a unitary transformation are physically equivalent
because a unitary transformation is defined as a linear norm-preserving
isomorphism of a Hilbert space onto itself\cite{DuS}.
More generally, however, two
theories are physically equivalent if they are related by an isometric
transformation\cite{DuS}:  a linear norm-preserving isomorphism of one
Hilbert space onto another.

A Hilbert space is a vector space together
with an inner product satisfying certain conditions.  Isometric
transformations allow the inner product to change under the transformation.
This freedom means that many non-unitary transformations define
physically equivalent theories.  Unitary equivalence has been
extensively studied from the standpoint of group representations\cite{Mac}.
Quantum canonical transformations give a somewhat different perspective.
This Letter will discuss the conditions
under which a quantum canonical transformation is an isometric
transformation.  The discussion is made for the quantum mechanical case,
but it is anticipated that it can be generalized to field theory.

To allow for time-dependent transformations, the time $q_0$ and its
conjugate momentum $p_0$ ($[q_0,p_0]=i$) are adjoined to the usual
canonical variables in phase space.  For notational convenience,
the collection of extended phase space
variables are denoted $(q,p)$, as if they were
one-dimensional.  The extension to higher dimensions is straightforward.

Since quantum canonical transformations will be defined outside of a
Hilbert space structure, two unfamiliar definitions are made.
First, the phase space variables $(q,p)$ are treated not as operators but
as elements of an associative algebra $\cal U$ containing all complex
functions of
$(q,p)$, consistent with the canonical commutation relations,
having Laurent expansions and their algebraic and
functional inverses (e.g. $(1+q)^{-1}$, $\ln q$).  Functions like
$p^{-n}$ are well-defined in this algebra.  Second, when acting on
functions on configuration space $\psi(q)$, the function $C(q,p)\in \U$
has the representation $\check{C}(\check{q},\check{p})$, where
$(\check{q}, \check{p})\equiv (q, -i\d_q)$.  Since $\check{C}$ is not in
general invertible as an operator, one takes the inverse in the algebra
and then represents it as an operator $(C^{-1})\check{\vb}$.  Operators
involving $(p^{-1})\check{\vb}$ are to be understood in the sense of
pseudo-differential operators\cite{Hor}.  For further discussion, see
Ref.~\cite{And}.

A quantum canonical transformation is defined\cite{And,Hei}
to be a change of the phase
space variables $(q,p)$, induced by a function $C(q,p)\in \U$,
\beqa
q &\mapsto& CqC^{-1} =q'(q,p), \\
p &\mapsto& CpC^{-1} =p'(q,p)  \non
\eeqa
which preserves the canonical commutation relations
\beq
[q,p]=i=[q'(q,p),p'(q,p)].
\eeq
There is a clear analogy with the definition of a classical canonical
transformation
as a change of the (classical) phase space variables which preserves the
Poisson bracket $\{q_c,p_c\}=1$.  Note that this definition does
not mention either Hilbert spaces or inner products.  The quantum
canonical transformations are neither unitary nor non-unitary.

For convenience, consider the case of non-relativistic quantum
mechanics.  The extension to relativistic quantum mechanics is immediate.
The Schrodinger operator corresponds to a function
$\sHa(q,p)=p_0+H(q_i,p_i,q_0)$ in $\U$.  The canonical transformation $C$
transforms $\sHa$ as
\beq
\label{sHtran}
\sHa'(q,p) = C\sHa(q,p)C^{-1}
=\sHa(Cq C^{-1},Cp C^{-1}).
\eeq
The solutions $\psi'$ of $\check{\sHa}'\psi'=0$ then induce solutions
$\psi$ of $\check{\sHa}\psi=0$
\beq
\label{wfinvtr}
\psi(q) =(C^{-1})\check{\vb} \psi'(q).
\eeq
Since no inner product has been specified, the transformation
$(C^{-1})\check{\vb}$
acts on all solutions of $\sHa'$, not merely the normalizable
ones.  There is some subtlety\cite{And} involving functions that lie
in the kernels of
$\check{C}$ or $(C^{-1})\check{\vb}$, but
it won't be important for physical equivalence since an isometric
transformation must be an isomorphism between the states of the
respective Hilbert spaces.

At this point, as we are interested in physical solutions, an inner product is
imposed on the solutions of $\sHa$
\beqa
\l \phi | \psi \r_{\mu} &\equiv& \l \phi | \mu(\check{q},\check{p})
| \psi \r_{1} \\
&=& \int d\Sigma\, \phi^*(q) \mu(\check{q},\check{p}) \psi(q), \nonumber
\eeqa
where the integration is over spatial configuration space.  Physical
solutions are those which are normalized to either unity or the delta function.
The ``measure density'' $\mu(\check{q},\check{p})$ may in general
be operator valued and may involve the temporal
variables.  This should not be surprising as the inner product for the
Klein-Gordon equation involves $\check{p}_0$.

When one makes a canonical transformation, in general the measure density
must transform to preserve the norm of states.
Given a canonical transformation $C$ from $\sHa$
to $\sHa'$ as in (\ref{sHtran}), the norm of states is preserved when
\beqa
\l \psi | \psi \r_{\mu} &=&
\l (C^{-1})\check{\vb}\,\psi' |\mu | (C^{-1})\check{\vb}\,\psi' \r_1 \non \\
&=& \l \psi' | (C^{-1})\check{\vb}\,\vb^\dag1\, \mu (C^{-1})\check{\vb}
|\psi' \r_1 \nonumber \\
&=& \l \psi' | \psi' \r_{\mu'}. \nonumber
\eeqa
The transformed measure density is
\beq
\label{mtran}
\mu'(\check{q},\check{p})=(C^{-1})\check{\vb}\,\vb^\dag1\,
\mu(\check{q},\check{p}) (C^{-1})\check{\vb},
\eeq
or, in $\U$,
\beq
\label{mutran}
\mu'(q,p)=C^{-1\, \dag1} C^{-1} \mu(Cq C^{-1},Cp C^{-1}).
\eeq
Here, $(C^{-1})\check{\vb}\,\vb^\dag1$ is the ``adjoint'' of
$(C^{-1})\check{\vb}$
in the trivial measure density, $\mu=1$. From (\ref{mutran}), one sees
that the measure transforms as a function on $\U$ multiplied by an
inhomogeneous
factor.

For functions of the spatial phase variables, the adjoint is computed by
taking the complex conjugate and integrating by parts (assuming boundary
terms to vanish---if they do not, the transformation is not a physical
equivalence). The adjoint cannot be taken for $p_0$ because the integration
measure does not involve $dq_0$. In some instances, a $p_0$-dependent
factor in a canonical transformation only changes the time-dependence of
the solutions in a unitary fashion. In this case, it can be allowed to act
directly on the wave function and its effect will cancel between the two
wave functions in the inner product. In other cases, canonical
transformations involving $p_0$ do not produce physically equivalent
systems.

If the Hilbert space of the transformed system $\sHa'$ has the measure given by
(\ref{mtran}) and is isomorphic to the original Hilbert space, the
canonical transformation is isometric.  The quantum theories defined by
$\sHa$ and $\sHa'$ and their Hilbert spaces are then physically
equivalent.

For isomorphisms of a Hilbert space onto itself, the measure density
does not change. If the measure density is purely a function of the
spatial coordinates, as it usually is in non-relativistic quantum
mechanics, one finds from (\ref{mutran}) the familiar condition for a
unitary transformation: $C^\dagger C=1$, where
$C^\dagger=\mu^{-1} C^\dag1 \mu$ is the adjoint
in the measure density $\mu$ of the Hilbert space.

An example will illustrate that (\ref{mtran}) is sometimes used implicitly
in common practice. Consider the canonical transformation $C=e^{q^2/2}$
from the harmonic oscillator $H=p^2 +q^2$, to the operator for the
Hermite polynomials $H'=p^2 +2iq p +1$. The harmonic oscillator energy
eigenstates $\psi_n(q)$
are given in terms of the Hermite polynomials $H_n(q)$
by
\beq
\psi_n(q)=e^{-q^2/2}H_n(q).
\eeq
The transformation is clearly an isomorphism, but the operator $\check{C}$
is not unitary because it is real. The harmonic
oscillator energy eigenstates are normalized in the trivial measure
$\mu=1$. By (\ref{mtran}), the transformation is an isometry and the
theories are physically equivalent if the Hermite
polynomials are normalized in the measure $\mu'=e^{-q^2}$. This is the
familiar result.

A canonical transformation can fail to be an isometry in two ways. First, the
transformation may not be an isomorphism. A canonical transformation or its
inverse that involves a finite-order differential operator will have a
non-trivial kernel. If the kernel of the transformation lies in one of the
Hilbert spaces, the transformation is not an isomorphism. This often
happens with raising and lowering or intertwining operators.

The simplest example is the canonical transformation given by the lowering
operator $p$ for the Hermite polynomials. This transformation annihilates
the ground state $\psi=1$ as it steps all the eigenfunctions down by one.
The mapping $p$ is not invertible on the space of Hermite polynomials. Such
a situation can sometimes be handled by speaking of partial isometries,
physical systems which are equivalent on a subsystem of states.

The second way that a canonical transformation can fail to be an isometry
is if the inner product in the transformed Hilbert space does not have the
measure (\ref{mtran}), so that the norms of states are not preserved. If
$C$ involves a finite-order differential operator, the norm-preserving
inner product will have an operator-valued measure density. In
non-relativistic quantum mechanics, measure densities are almost
exclusively coordinate-valued. If one does not use the norm-preserving
inner product, the normalization of states must have changed during the
transformation.

The simplest example is again given using the lowering operator for the
Hermite polynomials. The measure density which preserves the norm of states
after the transformation $\psi=p\psi'$ is $\mu'=pe^{-q^2}p$. If one uses
the measure density ${\tilde\mu}^\prime=e^{-q^2}$ appropriate to Hermite
polynomials, one finds an operator $e^{q^2}pe^{-q^2}p$ acting on one of the
states in the inner product. The Hermite polynomials are eigenfunctions of
this operator, so the form of the inner product is the same as it was
originally but with a state-dependent renormalization factor (the
eigenvalue). The norm of the states has changed. This is familiar because
the recursion operator between normalized $H_n$ depends on $n$. This
example is generic.

As a final example illustrating the broader application of isometric
canonical transformations to
physical equivalence than of unitary transformations,
consider two harmonic oscillators of different
frequency, $\sHa=p_0+p^2 +\w^2 q^2$ and $\sHa'=p_0 +p^2 +\w^{\prime\,2}
q^2$.  Perhaps unexpectedly, these are physically equivalent.  Their
spectra are different, so there can be no unitary transformation between
them.  The action of a canonical transformation $C$ in (\ref{sHtran})
on the function $\sHa$ corresponding to the Schrodinger
operator must be slightly generalized.  Solutions of $\sHa'$ are still
mapped to solutions of $\sHa$ if
\beq
\label{gsHtran}
\sHa'= D C\sHa C^{-1}
\eeq
where $D$ is taken in the present instance to be a constant, but could
more generally be a function of $q$ or any operator with a trivial kernel.
The canonical transformation
\beq
C=e^{i\ln(\w'/\w) q_0 p_0} e^{{i\over 2} \ln(\w'/\w) q p}
\eeq
relates the two harmonic oscillators through (\ref{gsHtran})
with the factor $D=\w'/\w$.  The harmonic oscillator energy eigenstates
of frequency $\w$ are given by
\beq
\psi'_n=\left( {\w' \over \pi} \right)^{1/4} e^{-\w' q^2/2}
H_n(\w^{\prime\,1/2} q) e^{-i(2n+1)\w' q_0}.
\eeq
The transformed solutions are
\beq
\psi_n=(C^{-1})\check{\vb}\,\psi'_n
=\left( {\w' \over \pi} \right)^{1/4} e^{-\w q^2/2} H_n(\w^{1/2} q)
e^{-i(2n+1)\w q_0}.
\eeq
Note that the normalization constant did not change.  This is because the
measure density transforms by (\ref{mtran}) to give
\beq
\mu'=e^{{i\over 2} \ln(\w/\w') pq}e^{{-i\over 2} \ln(\w/\w') q p}=
\left( {\w \over \w'} \right)^{1/2},
\eeq
so that the normalization of states is preserved. (In computing the
transformed measure density, the $p_0$ dependent part of $C$ is allowed to
act on the states to change their time-dependence, and only the adjoint of
the $p$ dependent term is taken.) Since the transformed measure is
independent of $n$, it can be absorbed into the definition of the wave
function to make the measure in the inner product trivial. That two
harmonic oscillators of different frequency should be physically equivalent
should not be counter-intuitive. Note that they are present in isolation so
that nothing is setting a time-scale. If one had a single system containing
two harmonic oscillators of different frequency, there would not be a
canonical transformation that would make both frequencies the same.

In summary, the definition of a general canonical transformation does not
involve the specification of a Hilbert space. Canonical transformations
therefore are not of themselves unitary or non-unitary. Two theories are
physically equivalent if the canonical transformation between them is an
isometric transformation. Upon specifying the initial and final Hilbert
spaces, a canonical transformation is an isometry if it is an isomorphism
and the measure densities in the inner products defining the Hilbert spaces
are related by (\ref{mtran}).

\vskip .5cm
I would like to thank J. Friedman and C.J. Isham for discussions of this
work.

\end{document}